\pdfoutput=1

\documentclass[showpacs,amsmath,amssymb,prd]{revtex4}
          
\usepackage{graphicx}
\usepackage{dcolumn}
\usepackage{bm}
\usepackage{epsfig}
\usepackage{amsmath}
\usepackage{color}

\newcommand{\be}{\begin{equation}}
\newcommand{\ee}{\end{equation}}
\newcommand{\ba}{\begin{eqnarray}}
\newcommand{\ea}{\end{eqnarray}}

\begin{document}
\input{epsf}

\title{The Galactic Center Origin of a Subset of IceCube Neutrino Events}

\author{Soebur Razzaque}

\email{srazzaque@uj.ac.za}

\affiliation{Department of Physics, University of Johannesburg, PO Box
  524, Auckland Park 2006, South Africa}

\begin{abstract} 
  The center of the Milkyway is a host to energetic phenomena across
  many electromagnetic wave-bands and now possibly of high-energy
  neutrinos. We show that 5 out of 21 IceCube shower-like events,
  including a PeV event, likely originated from the Galactic Center
  region.  Hard spectrum and flux inferred from these events are
  inconsistent with atmospheric neutrinos.  The flux of these
  neutrinos is consistent with an extrapolation of the gamma-ray flux
  measured by Fermi-LAT from the inner Galactic region.  This
  indicates a common hadronic origin of both, powered by supernovae.
  Three other shower-like events are spatially correlated with the
  Fermi bubbles, originating from the Galactic Center activity, within
  the uncertainty of reconstructing their arrival directions.  Origin
  of the other neutrino events, including 7 track-like events, is
  still elusive.
\end{abstract}


\date{\today}
\maketitle


The IceCube Collaboration has recently announced detection of 26
neutrino events with energies in the $\sim 30$--250 TeV range
\cite{icevents}, in addition to the two events announced earlier with
$\sim 1$ PeV energy each \cite{Aartsen:2013bka}.  Among these 28
contained vertex events, for which electromagnetic energy deposition
in the detector from $\nu_l\to l + X$; $l=e, \mu, \tau$ interactions
can be reliably reconstructed, 21 are shower-like events (including
the two PeV events) without visible muon tracks and 7 are muon
track-like events.  These events, collected over 662 days livetime,
constitute a $4.3\sigma$ signal over an expected background of
$10.6^{+4.5}_{-3.9}$ events from the atmospheric neutrino flux models
\cite{icevents}.  Although an atmospheric origin of these $\nu$'s
cannot be ruled-out completely, this is most likely the first
statistically significant signal of high-energy astrophysical
neutrinos \cite{Laha:2013lka, Lipari:2013taa}.

\begin{figure}[ht]
\includegraphics[width=6.in]{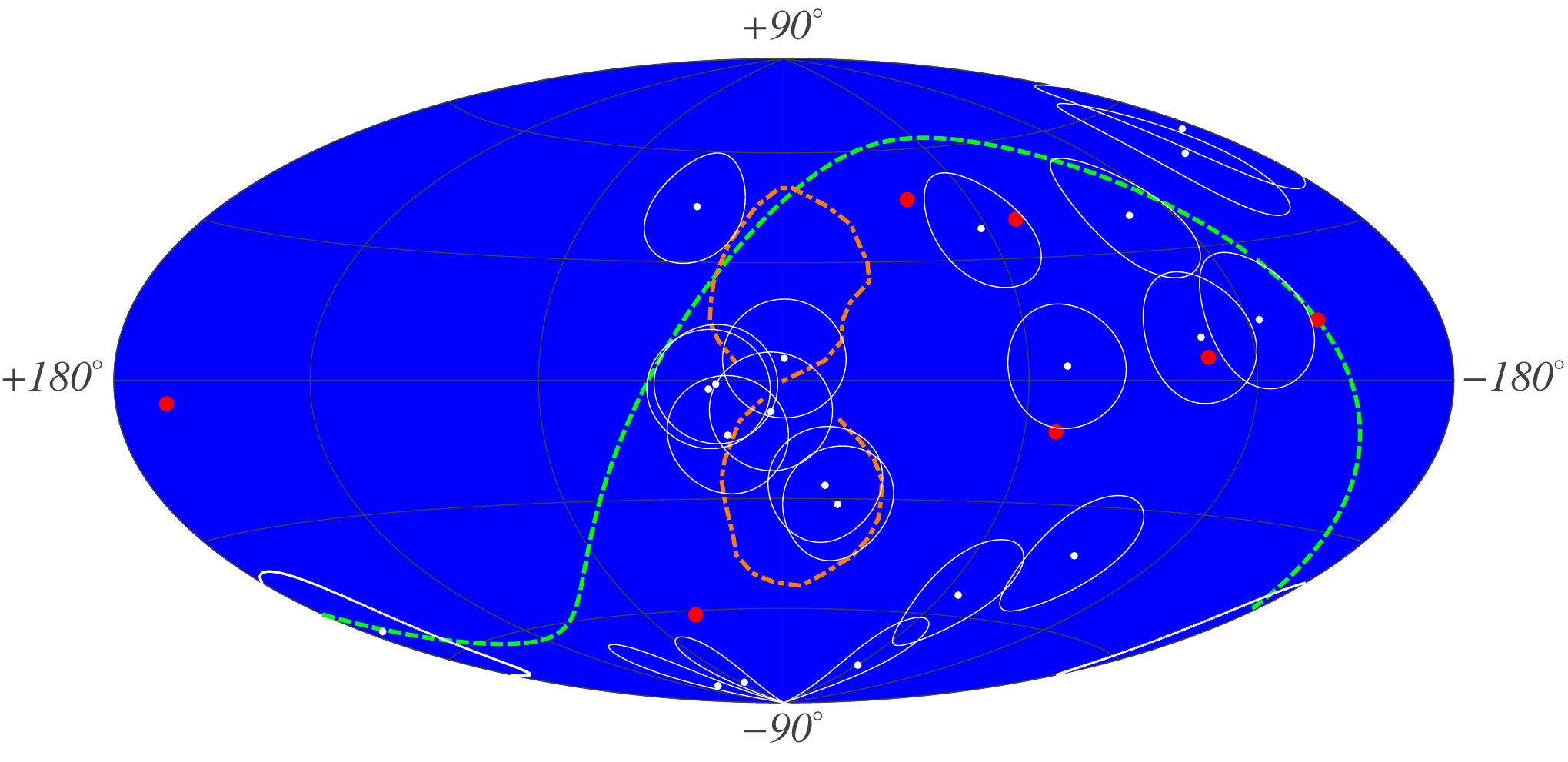}
\caption{IceCube neutrino events in Galactic coordinates.  The 21
  shower-like events are shown with $15^\circ$ error circles around
  the approximate positions (small white points) reported by IceCube
  \cite{icevents}.  The 7 track-like events are shown as larger red
  points. Also shown are the boundaries of the Fermi bubbles
  (dot-dashed line) and the Equatorial plane (dashed line).}
\label{fig:icevents}
\end{figure}

Figure \ref{fig:icevents} shows the arrival directions of the 28
IceCube neutrino events, adapted from the skymap in
Ref.~\cite{icevents}, in Galactic coordinates.  A $15^\circ$ error
circle has been drawn around each of the shower-like events, to
reflect the uncertainty in reconstructing the arrival directions of
these kind of events~\footnote{N.~Whitehorn, private communication}.
The arrival directions of the track-like events has an uncertainty of
the order of a few degree.  It is interesting to note that the largest
concentration of events (5 shower-like) is at or near the Galactic
Center, within uncertainties of their reconstructed directions, and an
additional 3 shower-like events have their arrival directions
consistent with the Fermi bubbles \cite{Su:2010qj}.  There is an
absence of any track-like events in this region.  Most of the
track-like events are out of the Galactic plane and at least 4 of them
are correlated with shower-like events in those regions.


The IceCube Collaboration has reported \cite{Aartsen:2013bka} the
average neutrino effective areas for the shower-like events, over
$4\pi$ solid angle and at 1~PeV, as $A_{e}\sim 5$~m$^2$, $A_{\mu}\sim
1$~m$^2$, $A_{\tau}\sim 2$~m$^2$, respectively for $\nu_e$, $\nu_\mu$
and $\nu_\tau$ ($\nu$ refers to $\nu + {\bar \nu}$ throughout). For
the two events at $\sim 1$~PeV each within $t_{live} = 615.9$ days
livetime, the required isotropic diffuse $\nu$-flux at $E = 1$~PeV is
\be
E^2 \Phi (\nu_e + \nu_\mu + \nu_\tau) \approx \frac{1}{4\pi}
\frac{2\times 10^6~{\rm GeV}}
{(A_e + A_\mu + A_\tau)t_{live}}
 \sim 3.7\times 10^{-8} ~{\rm GeV}~{\rm sr}^{-1}
~{\rm cm}^{-2}~{\rm s}^{-1},
\label{diff_flux}
\ee
as also calculated in Ref.~\cite{Aartsen:2013bka}. However, this flux
could be $\sim 20\%$ lower when the effective area for muon track-like
events are included \footnote{L.~Anchordoqui, private communication}.
The effective areas are a factor of $\sim 3$ smaller at 100~TeV, as
can be found by comparing them for the 22 string IceCube confguration
\cite{Abbasi:2011ui}.  Among the 5 shower-like events which are
spatially consistent with the Galactic Center region, 1 event has
$\sim 1.1$~PeV energy while the rest have energies in the $\sim
30$--250 TeV range.  Note that the reported energy is
electromagnetic-equivalent energy deposited in showers
\cite{icevents}.  The actual neutrino energy is higher.  Both
neutral-current (NC) and charge-current (CC) interactions of neutrinos
can produce shower events, the deposited energy in showers, however,
is significantly different between these two interaction channels.
While for CC the shower energy (for events without visible muon
tracks) is almost equivalent to the neutrino energy, for NC the shower
energy is $\approx 40\%$-$25\%$ of the neutrino energy in the 10
TeV-10 PeV range \cite{Gandhi:1995tf}.  Bearing this uncertainty of
neutrino energy reconstruction in mind, we assume a common 100 TeV
energy for these 4 events.  Assuming all 5 of these events originated
from within a $8^\circ$ circular region around the Galactic Center
(solid angle $\Omega_{GC} = 2\pi (1-\cos 8^\circ) = 0.06$~sr), given
$\sim 15^\circ$ uncertainty in reconstructing the arrival directions
of these events (see Fig.~\ref{fig:icevents}), we calculate $\nu$
fluxes at 100~TeV and 1~PeV from this region as
\ba
E^2 \Phi (\nu_e + \nu_\mu + \nu_\tau) 
\sim \begin{cases} 
1.3\times 10^{-9}~{\rm GeV} ~{\rm cm}^{-2}~{\rm s}^{-1} 
~;~ (100~{\rm TeV}) \cr
1.1\times 10^{-9}~{\rm GeV} ~{\rm cm}^{-2}~{\rm s}^{-1}
~;~ (1~{\rm PeV}), 
\label{GC_fluxes}
\end{cases}
\ea
following Eq.~(\ref{diff_flux}) but with $t_{live} = 662$ days and by
multiplying with $\Omega_{GC}$. This corresponds to a hard, $\Phi
\propto E^{-\Gamma}$ with $\Gamma\sim 2$, $\nu$-flux from the Galactic
Center region. Given the $7:1$ ratio of the effective areas for the
$\nu_e + \nu_\tau : \nu_\mu$ shower-like events, the ratio of the
$\nu_e + \nu_\tau : \nu_\mu $ fluxes from the Galactic Center region
can be nearly equal and yet no detection of any $\nu_\mu$ events so
far.  Including the energy-dependent effective area for muon
track-like events changes the ratio somewhat but the conclusion about
the nearly equal flavor ratio does not change.  As for comparisons,
the expected conventional atmospheric $\nu$-flux \cite{Honda:2006qj}
from the Galactic Center region \footnote{Galactic Center's $(RA, Dec)
  = (266.4^\circ, -28.9^\circ)$ implies a zenith angle of $\theta_z =
  120.6^\circ$ for IceCube} within $\Omega_{GC}$ is
\be
E^2 \Phi (\nu_\mu) = 5.0\times 10^{-11} 
\left( E/{\rm PeV}\right)^{-1.6}
~{\rm GeV} ~{\rm cm}^{-2}~{\rm s}^{-1},
\label{atm_flux}
\ee
much smaller than the PeV flux in Eq.~(\ref{GC_fluxes}). At 100 TeV
the atmospheric $\nu_\mu$ flux is comparable to the flux in
Eq.~(\ref{GC_fluxes}).  However, note that the atmospheric $\nu_e$
flux is steeper and at least a factor of 10 smaller than the $\nu_\mu$
flux in Eq.~(\ref{atm_flux}) in this energy range.  Moreover, neutrino
oscillation does not produce significant $\nu_\tau$ flux at these
energies. Therefore atmospheric neutrino fluxes are inconsistent with
the nearly equal or higher $\nu_e + \nu_\tau : \nu_\mu$ flux ratio
expected from the 5 shower-like events detected by IceCube from the
Galactic Center region.

Interestingly, the Fermi-LAT measures \cite{Ackermann+12} a 100 GeV
$\gamma$-ray flux of $\sim 3.5\times
10^{-3}$~MeV~cm$^{-2}$~s$^{-1}$~sr$^{-1}$ from the inner Galactic
region ($|l| < 80^\circ$ and $|b| < 8^\circ$).  This flux, within
$\Omega_{GC}$, is $\sim 2.1\times 10^{-7}$~GeV~cm$^{-2}$~s$^{-1}$
which is far above the $\nu$-fluxes in Eq.~(\ref{GC_fluxes}).  An
extrapolation of the Fermi-LAT flux in the 0.1--1~PeV range is highly
uncertain, due to non-power-law nature of this flux.  Moreover, both
$\pi^0$ decay and inverse Compton emissions contribute to this flux at
high energies \cite{Ackermann+12}.  Extrapolating the $\pi^0$ decay
$\gamma$-ray flux of $\sim 10^{-3}$~MeV~cm$^{-2}$~s$^{-1}$~sr$^{-1}$
at 100~GeV, with $E^{-2.3}$ spectrum, to 100~TeV and 1~PeV gives a
flux of $\sim 7.6\times 10^{-9}$~GeV~cm$^{-2}$~s$^{-1}$ and $\sim
3.8\times 10^{-9}$~GeV~cm$^{-2}$~s$^{-1}$, respectively, within
$\Omega_{GC}$.  These fluxes are comfortably above the $\nu$-fluxes in
Eq.~(\ref{GC_fluxes}) derived from the IceCube data, hinting that the
same cosmic-ray interactions produce the observed $\gamma$ rays and
$\nu$'s.

Indeed, the Galactic Center has been considered as a plausible $\nu$
source in the past \cite{Stecker:1978ah}. Estimates of the Galactic
Center $\nu$-flux from $pp$ interactions, with a steep $E^{-2.7}$
cosmic-ray spectrum based on observations on the Earth, resulted in
steep $\nu$ spectrum similar to the atmospheric $\nu$-flux.  This gave
a Galactic Center $\nu$-flux estimate lower than the atmospheric
$\nu$-flux models, which is inconsistent with IceCube observations as
we have discussed earlier.  We discuss below a likely origin of the
hard-spectrum $\nu$-flux in Eq.~(\ref{GC_fluxes}) from the Galactic
Center region.

The inner Galaxy is a host to active star-formation activity
\cite{Serabyn+1996, Figer:2003tu}. The $\sim 1.3\times 10^{39}$~erg/s
in cosmic-ray power at the Galactic Center \cite{Crocker:2010dg},
inferred from the supernovae rate, is far above the required
bolometric $\nu$-luminosity of $\sim 4.1\times 10^{34}$~erg/s within
$\Omega_{GC}$, over one decade in energy, derived from
Eq.~(\ref{GC_fluxes}).  The acceleration of cosmic-rays in the
supernovae takes place over the Sedov time scale of a few thousand
year.  These cosmic-rays are trapped in the Galactic Center region and
lose energy before escaping.  The cooling time scale for the
cosmic-ray protons due to $pp$ interactions, with an average
cross-section $\sigma_{pp} = 60$~mb and inelasticity $k_{pp}=0.5$, is
\be
t_{pp} = (\sigma_{pp} n_{H} k_{pp})^{-1}
\sim  1.2\times 10^7 (n_{H}/3~{\rm cm}^{-3})^{-1} ~{\rm year},
\label{pp_cool}
\ee
where $n_{H} \sim 3$~cm$^{-3}$ is the average gas density within the
central 1~kpc of the Galactic Center, enclosing a total gas mass of
$M_H \sim 6\times 10^7 M_\odot$ \cite{Launhardt:2002tx}. The escape
time scale for the cosmic rays from within $8^\circ$ of the Galactic
Center region, with radius $R\sim 1.2$~kpc, in the Bohm diffusion
limit is
\be
t_{esc} = \frac{3}{2} \frac{m_p^2}{E_p} \frac{B_G}{B_{cr}}R^2
\sim 5.4\times 10^6 \left(\frac{E_p}{10\,\rm PeV}\right)^{-1} 
\left( \frac{B_G}{12~{\mu\rm G}} \right)
\left( \frac{R}{\rm kpc} \right)^2 ~{\rm year},
\ee
where $B_G \sim 12~\mu$G is the magnetic field in the Galactic Center
region and $B_{cr} = 1.488\times 10^{20}$~G is the critical magnetic
field for protons. A similar or longer escape time scale for cosmic
rays with the maximum 10~PeV or lower energy, as compared to the $pp$
cooling time scale in Eq.~(\ref{pp_cool}) results in a hard cosmic-ray
spectrum, essentially the same as the injected spectrum from the
supernovae, in the Galactic Center region.  The $\nu$ spectrum from
the $pp$ interactions should therefore be hard too, as indicated by
Eq.~(\ref{GC_fluxes}) derived from IceCube data. Note that this is the
same $t_{acc} < t_{pp} \lesssim t_{esc}$ argument for the Fermi bubble
$\gamma$-ray and $\nu$ spectra \cite{Crocker:2010dg,
  Lunardini:2011br}, which are rather hard ($\Gamma \sim 2$).

The 3 shower-like $\nu$ events which are spatially correlated with the
Fermi bubbles at relatively high Galactic latitudes, have energies in
the $\sim 30$--250 TeV range.  The $\gamma$-ray flux from the Fermi
bubbles is also contamination-free and more reliable at high Galactic
latitudes \cite{Su:2010qj}.  If hadronic mechanism is responsible for
the bubble $\gamma$-ray flux, the 3 shower-like $\nu$ events could
originate from the same cosmic-ray interactions.  An order of
magnitude estimate follows from the Fermi bubble $\nu$-fluxes
predicted in Ref.~\cite{Lunardini:2011br}.  At 100 TeV the $\nu$-flux
on the Earth, with equal flavor ratios after oscillation, is $E^2 \Phi
(\nu_e + \nu_\tau) \sim 2\times
10^{-7}$~GeV~sr$^{-1}$~cm$^{-2}$~s$^{-1}$ for the 10 PeV cosmic-ray
energy cutoff case.  Considering only half (at high Galactic
latitudes) of the Fermi bubble solid angle, $\Omega_{FB/2} =
0.404$~sr, we calculate the number of 100 TeV shower-like $\nu$ events
as
\be
N_{FB/2} 
\sim E\Phi (\nu_e + \nu_\tau) \Omega_{FB/2} (A_e+A_\tau) t_{live}
\sim 1.1,
\ee
which is fewer than the number IceCube has detected.  The 100 TeV
$\nu$-flux from the Fermi bubbles could be a factor $\sim 2$--3 higher
in principle, given the uncertainty on the $\gamma$-ray data and/or if
cosmic-ray cutoff energy is $> 10$~PeV.  A more detailed calculation
of the $\nu$ events from the Fermi bubbles for the full IceCube will
be presented elsewhere \cite{Lunardini:2013, Ahlers:2013}.

The origin of the other 20 IceCube $\nu$ events, including a PeV
shower-like event and all 7 track-like events is less clear. It was
pointed out in Ref.~\cite{Fox:2013oza} that up to 2 events could
originate from the Galactic TeV unidentified sources. A pulsar wind
nebula HESS J1825-137 and a supernova remnant G 0.9+0.1 near the
Galactic Center have been considered as the origin of the IceCube
$\nu$ events as well \cite{Neronov:2013lza}. A general scenario of
Galactic $\gtrsim 10$~PeV cosmic-ray interactions to produce IceCube
PeV events \cite{Gupta:2013xfa} and plausible spectra of IceCube $\nu$
events as originating from Galactic cosmic-rays
\cite{Anchordoqui:2013qsi} have been considered as well.  Some of the
reported $\nu$ events, specially below PeV energy, could originate
from extragactic sources, such as jets inside GRB/hypernova
progenitors \cite{Razzaque:2003uv, Razzaque:2004yv, Murase:2013ffa}.
Ultrahigh-energy cosmic-ray interactions in the jets of AGNs
\cite{Stecker:2013fxa} and GRBs \cite{Razzaque:2013dsa}, and/or during
propagation in the CMB \cite{Kalashev:2013vba} could produce the
observed PeV $\nu$ event(s).  Finally, PeV scale dark-matter decay has
been considered to explain IceCube $\nu$ events as well
\cite{Esmaili:2013gha}.  More data in future will be able to shed
light on possible Galactic and extragalactic sources.

In summary, we have shown that cosmic-rays accelerated by supernovae
in the Galactic Center region is likely the origin of 5 shower-like
$\nu$ events detected by IceCube in the $\sim 0.1$-1 PeV range from
this region. An additional 3 shower-like events may originate from the
Fermi bubbles at high Galactic latitude.  This scenario is consistent
with hard spectrum $\gamma$-ray flux measured by Fermi-LAT from the
inner Galaxy and from the Fermi bubbles, pointing to a common origin of
$\gamma$ rays and $\nu$'s from the Galactic cosmic-ray interactions.

\acknowledgments

I would like to thank Aya Ishihara, Cecilia Lunardini and Nathan
Whitehorn for discussion and for providing useful information. I also
thank Markus Ahlers, Luis Anchordoqui, Arman Esmaili and Francis
Halzen for comments.

\end{document}